\documentclass[12pt]{article}
\usepackage{latexsym}
\usepackage{amsmath}
\usepackage{amssymb}
\usepackage{amsfonts}
\usepackage{graphicx}
\usepackage{epsfig}
\usepackage{array}

\newtheorem{ExampleDef}{Example}[section]

\newcommand{\Example}[3]{
  \begin{list}{}{
      \setlength{\leftmargin}{1em}}     
    \item                               
    \small                              
    \begin{ExampleDef} \rm              
      {\bf \hspace{-1ex}: #1}           
      #2                                
      \hfill {\large \boldmath $\Box$}  
      \label{ex:#3}                      
    \end{ExampleDef}
  \end{list}}

\begin{document}
\begin{center}
{\Large {\bf $K$-Winners-Take-All Computation\\
with Neural Oscillators\par}}
\vspace{1.0em}
{\large Wei Wang and Jean-Jacques E. Slotine\footnote{To whom correspondence 
should be addressed.} \par} 
{Nonlinear Systems Laboratory \\
Massachusetts Institute of Technology \\
Cambridge, Massachusetts, 02139, USA 
\\ wangwei@mit.edu, \ jjs@mit.edu 
\par}
\vspace{2em}
\end{center}

\begin{abstract}
Artificial spike-based computation, inspired by models of computation
in the central nervous system, may present significant performance
advantages over traditional methods for specific types of large scale
problems. This paper describes very simple network architectures for
$k$-winners-take-all and soft-winner-take-all computation using neural
oscillators. Fast convergence is achieved from arbitrary initial
conditions, which makes the networks particularly suitable to track
time-varying inputs.
\end{abstract}


%
\section{Introduction} \label{sec:introduction}
The discovery of synchronized oscillations in the visual cortex and
other brain regions has triggered significant research in artificial
spike-based computation~\cite{hopfield03, gerstner, gray, 
jin02, llinas03, llinas98, llinas02, maass03, singer, thorpe01,
malsburg95, deliang02}. While neurons in the central nervous
system are about six orders of magnitude "slower" than silicon-based
elements, in both elementary computation time and signal transmission
speed, their performance in networks often compares very favorably
with their artificial counterparts even when reaction speed is
concerned. In a sense, evolution may have been forced to develop
extremely efficient computational schemes given available hardware
limitations.

In a recent paper~\cite{wei03-1}, we proposed new models for two
common instances of such neural computation, winner-take-all and
coincidence detection, featuring fast convergence and $O(n)$ network
complexity.  We saw that both computations could be achieved using a
similar architecture, using global feedback inhibition in the first
case, and global excitation in the second.  In this paper, we further
extend this computational architecture to $k$-winners-take-all and
soft-winner-take-all.

Fast Winner-take-all (WTA) computation in~\cite{wei03-1} is based on
the FitzHugh-Nagumo model, a well-known simplified version of the
classical Hodgkin-Huxley model. Compared to previous WTA
networks~\cite{arbib, fang, grossberg73, jin02,lazzaro, yuille89}, it
has significant computational advantages. The network's initial states
can be set arbitrarily, and convergence is guaranteed in at most two
spiking periods, with a high computation resolution. The network's
complexity is linear in the number of inputs and its size can be
adjusted at any time during the computation. As this paper shows, by
modifying the starting point of the global inhibitory neuron's
charging mode, $k$-Winners-Take-All ($k$-WTA) can be computed instead.
By running the charging mode independently, soft-Winner-Take-All
(soft-WTA) can be computed. Both extensions inherit the advantages of
the original WTA network.

After a brief review of the basic WTA network in
section~\ref{sec:wta}, $k$-WTA computation and soft-WTA computation
are studied in Sections~\ref{sec:kwta} and~\ref{sec:soft}.  Brief
concluding remarks are offered in Section~\ref{sec:conclusion}.

\section{Winner-Take-All Network} \label{sec:wta}
The WTA network in~\cite{wei03-1} is based
on the FitzHugh-Nagumo (FN) model~\cite{fitzhugh,nagumo,murray}:
\begin{equation*} \label{eq:f-n}
\begin{cases}  
  \ \dot{v} = v(\alpha - v)(v-1)-w+I  \\  
  \ \dot{w} = \beta v - \gamma w
\end{cases}  
\end{equation*}
For appropriate parameter choices, there exists
a unique equilibrium point for any given value of $I$, which
is stable except for a finite range $\ I_l \le I \le I_h\ $ 
where the system tends to a limit cycle.  The steady-state 
value of $v$ at the stable equilibrium point increases
with $I$.

\begin{figure}[h]
\begin{center}
\epsfig{figure=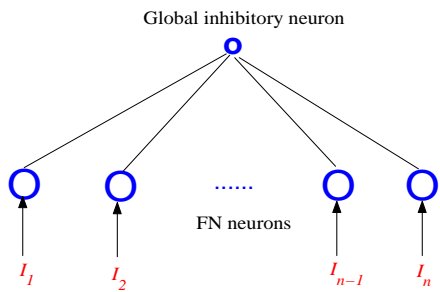,height=40mm,width=60mm}
\caption{Diagram of the WTA network. There are $n$ FN neurons
receiving external inputs and a global inhibitory neuron 
monitoring the whole network.}
\label{fig:structure}
\end{center}
\end{figure}

The network structure is illustrated in Figure~\ref{fig:structure}
where $n$ FN neurons receive external stimulating inputs $I_i$ and
a global inhibition $z$. The dynamics of the FN neurons 
($i=1, \ldots, n$) are
\begin{equation*} \label{eq:fn-in-wta}
\begin{cases}  
  \ \dot{v}_i = v_i (\alpha - v_i) (v_i-1)-w_i + I_i - z  \\  
  \ \dot{w}_i = \beta v_i - \gamma w_i
\end{cases}  
\end{equation*}
The dynamics of the global inhibitory neuron is
\begin{equation} \label{eq:inhibition-in-wta}
\dot{z} = 
\begin{cases}  
  \  - k_c \ (z - z_0)  \ \ \ \ \mathrm{charging}\ \mathrm{mode} \\  
  \  - k_d \ z   \ \ \ \ \ \ \ \ \ \ \ \ \mathrm{discharging}\ \mathrm{mode}
\end{cases}  
\end{equation}
which starts charging if there is any FN neuron spiking in the network
(i.e., if $\exists i$, $v_i \ge v_0$ for some given threshold $v_0$)
and switches to discharging if the state $z$ is saturated. With a fast
charging rate $k_c$ and a slow discharging rate $k_d$, the network
computes the largest input (corresponding to the only spiking FN
neuron) in at most two periods. Initial conditions can be set
arbitrarily and the computation resolution is very high. Detailed
analysis and discussions can be found in~\cite{wei03-1}.

\section{$K$-Winners-Take-All Network} \label{sec:kwta}
$K$-WTA, a common variation of WTA computation where the output
indicates for each neuron whether its input is among the $k$ largest,
has been studied in such fields as competitive learning, pattern
recognition and pattern classification~\cite{badel, fukai, urahama, 
wolfe, juicheng}.  As Maass argued in~\cite{maass00-1}, in principle a
$k$-WTA network can replace a two-layer threshold circuit to perform
most standard nonlinear computational operations.

Most $k$-WTA studies are based on steady-state stability
analysis. Many models define the winners as the neurons with the
largest initial states~\cite{majani, wolfe} or require initial
conditions to be set precisely~\cite{juicheng}, making the networks
not well suited to time-varying inputs.  Others adopt particular
design methodologies~\cite{perfetti, seiler} but the network size or
the number of winners is limited. $K$-WTA is also implemented in
analog VLSI circuits~\cite{urahama}, which extend the elegant WTA
model in~\cite{lazzaro} but inherit its low resolution limit as well.

The neural network described in Section~\ref{sec:wta} can be easily
extended to $k$-WTA computation, where an FN neuron spikes if and only
if its input is among the $k$ largest. Indeed, as the global
inhibition force decreases, the FN neurons enter the oscillation
region {\it rank-ordered} by their inputs. Thus, while for WTA
computation, the global inhibition neuron is charged after the first
arrival, for $k$-WTA computation the charging moment is simply
modified to capture the $k^{\rm th}$ arrival instead.

To this effect, we augment the dynamics of the FN neuron with an
additional state variable $u_i$ (for simplicity, we shall still call
FN neuron such a generalized element)
\begin{equation*} \label{eq:fn-in-kwta}
\begin{cases}  
  \ \dot{v}_i = v_i (\alpha - v_i) (v_i-1)-w_i + I_i - u_i - z  \\  
  \ \dot{w}_i = \beta v_i - \gamma w_i \\
  \ \dot{u}_i = k_u \ (\zeta_i u_0 - u_i) 
\end{cases}  
\end{equation*}
where $u_0$ is a constant saturation value and $k_u$ the
charging/discharging rate.  The variable $\zeta_i$ takes two values,
namely it switches to $0$ whenever $z$ approaches a saturation value
$z_0$, else it switches to $1$ if $v_i$ exceeds a given threshold
$v_0$. This make the dynamics of $u_i$ a local self-inhibition, which
starts charging if the basic FN neuron spikes and discharges whenever
the global inhibitory neuron spikes. Note that the specific form of
the dynamics of $u_i$ can be more general, as long as the value of
$u_i$ varies between $0$ and $u_0$, and the transition periods are
very fast (which is satisfied here by choosing a large $k_u$).

The dynamics of the global inhibitory neuron is the same
as~(\ref{eq:inhibition-in-wta}), except that we start its charging
mode if any $k$ FN neurons in the network spike. Such a moment can be
captured by determining that $\ \sum_{i=1}^n u_i\ $ approaches $k
u_0$. Thus, if any FN neuron spikes, it excites only the corresponding
local inhibitory portion but has no effect on the rest of the
network. If there are $k$ local inhibitions turned on, the global
inhibitory neuron is charged, which then releases all the local
inhibitions and starts a new period.

Compared to the WTA network in Section~\ref{sec:wta}, the basic
principle underlying the $k$-WTA network described above is the same,
exploiting the simple properties of the FN model.  Thus, most of the
computational advantages of the WTA network~\cite{wei03-1} are
inherited by the $k$-WTA extension.  In particular
\begin{itemize}
\item The initial conditions of the network can be set arbitrarily. 

\item With appropriate parameters, the computation can be completed at
most in two periods, where the first period is affected by the initial
conditions but the $k$ spiking neurons during the following periods
are guaranteed to be those with the largest inputs. If the initial
inhibitions are strong, the computation is completed in one period.

\item Since initial conditions are immaterial and the computation
speed is very fast, the $k$-WTA network is able to track time-varying
inputs. Moreover, since the network complexity is $O(n)$, individual
FN neurons can be added or removed at any time during the computation.

\item The inputs $I_i$ should be lower-bounded by $I_l$, the lower
threshold of the FN oscillation region. It should also be
upper-bounded to set inhibition saturations, although the upper bound
value is not restricted.

\item  The computation resolution also follows that of the WTA network. It can 
be improved by decreasing the discharging rate $k_d$, as well as the relaxation
time of the FN neurons. 

\item FN neurons receiving equal inputs behave identically, which
means that the $k$-WTA computation may generate more than $k$ winners
in this particular case.

\end{itemize}

\begin{figure}[p]
\begin{center}
\epsfig{figure=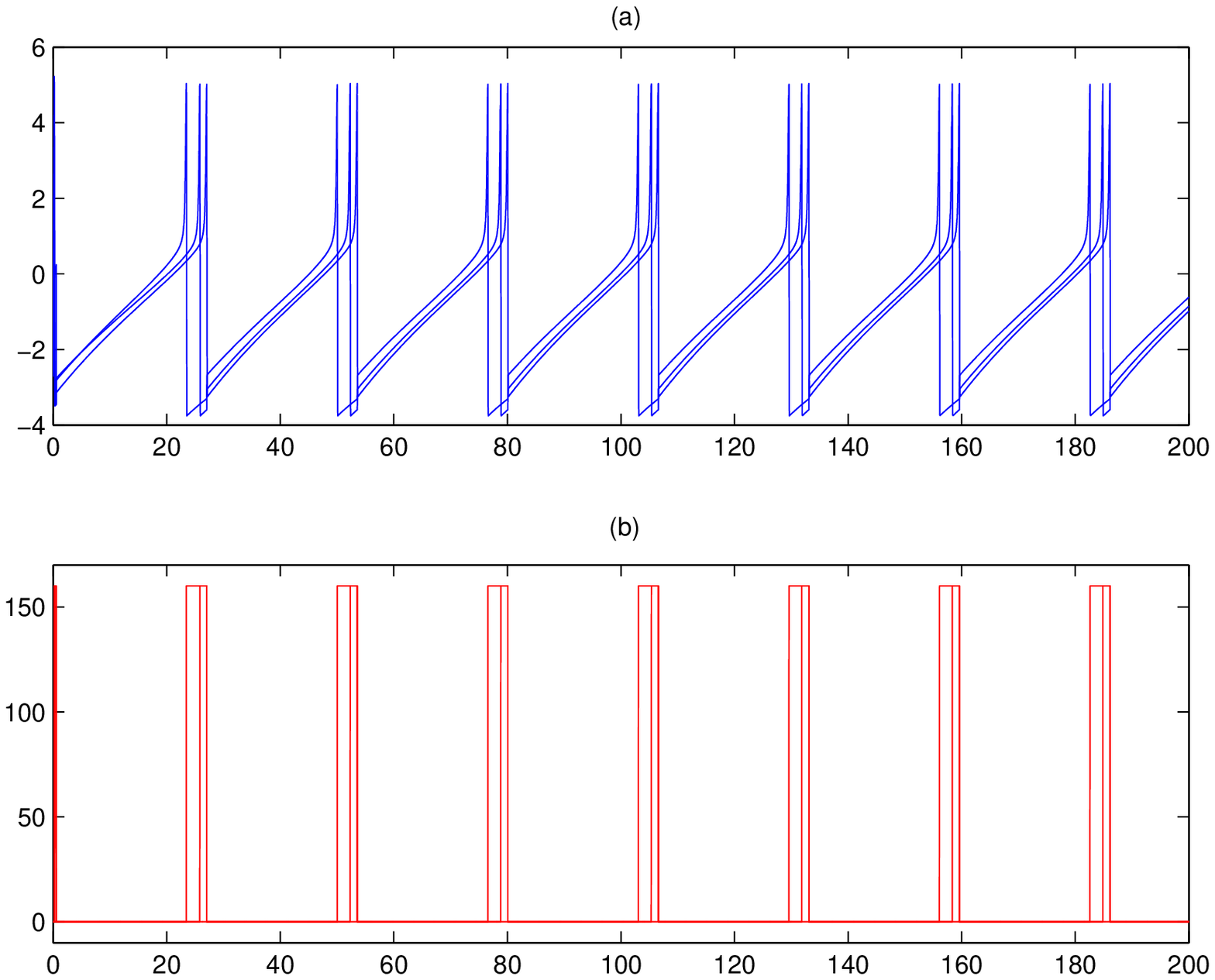,height=60mm,width=110mm}
\epsfig{figure=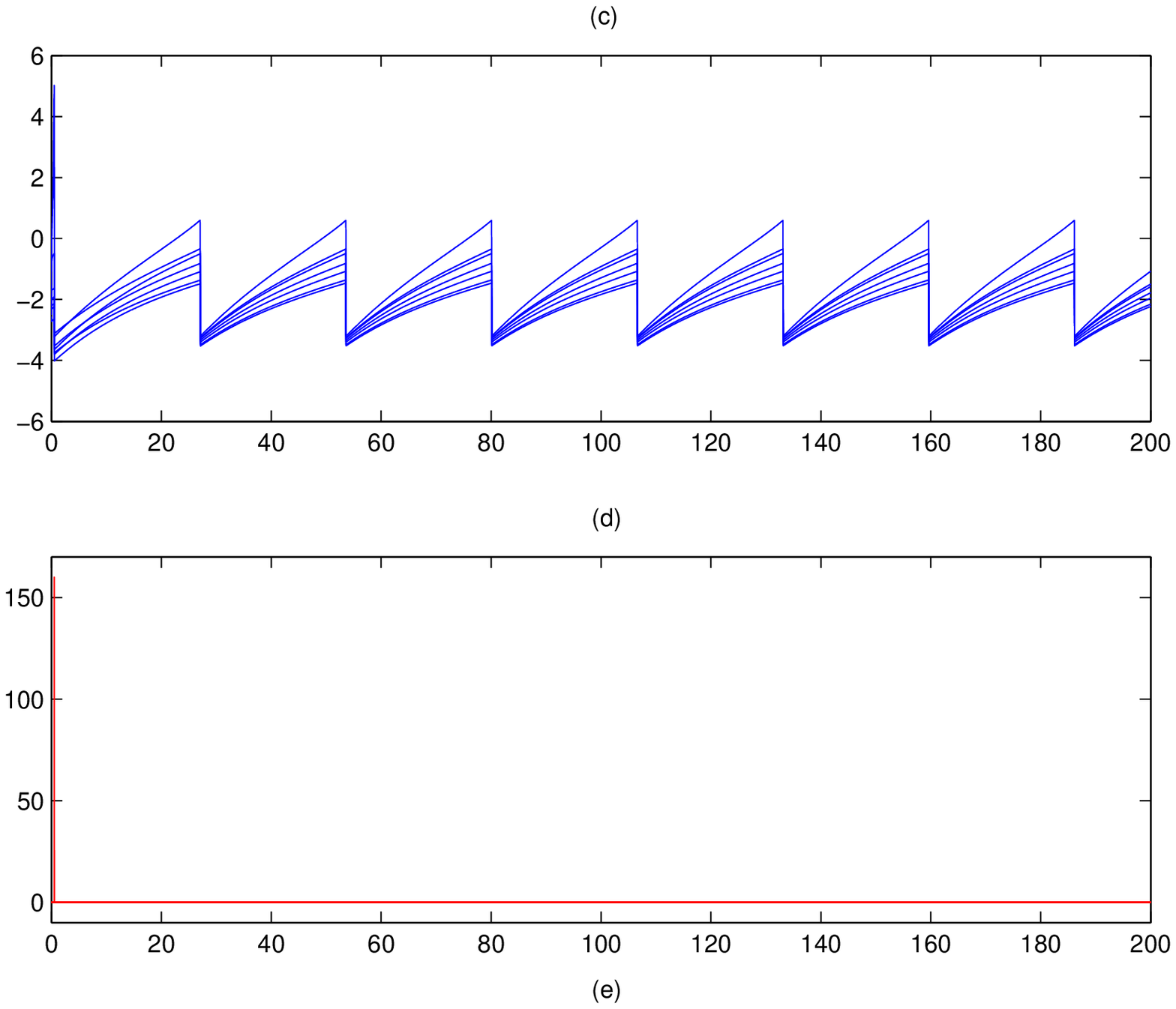,height=60mm,width=110mm}
\epsfig{figure=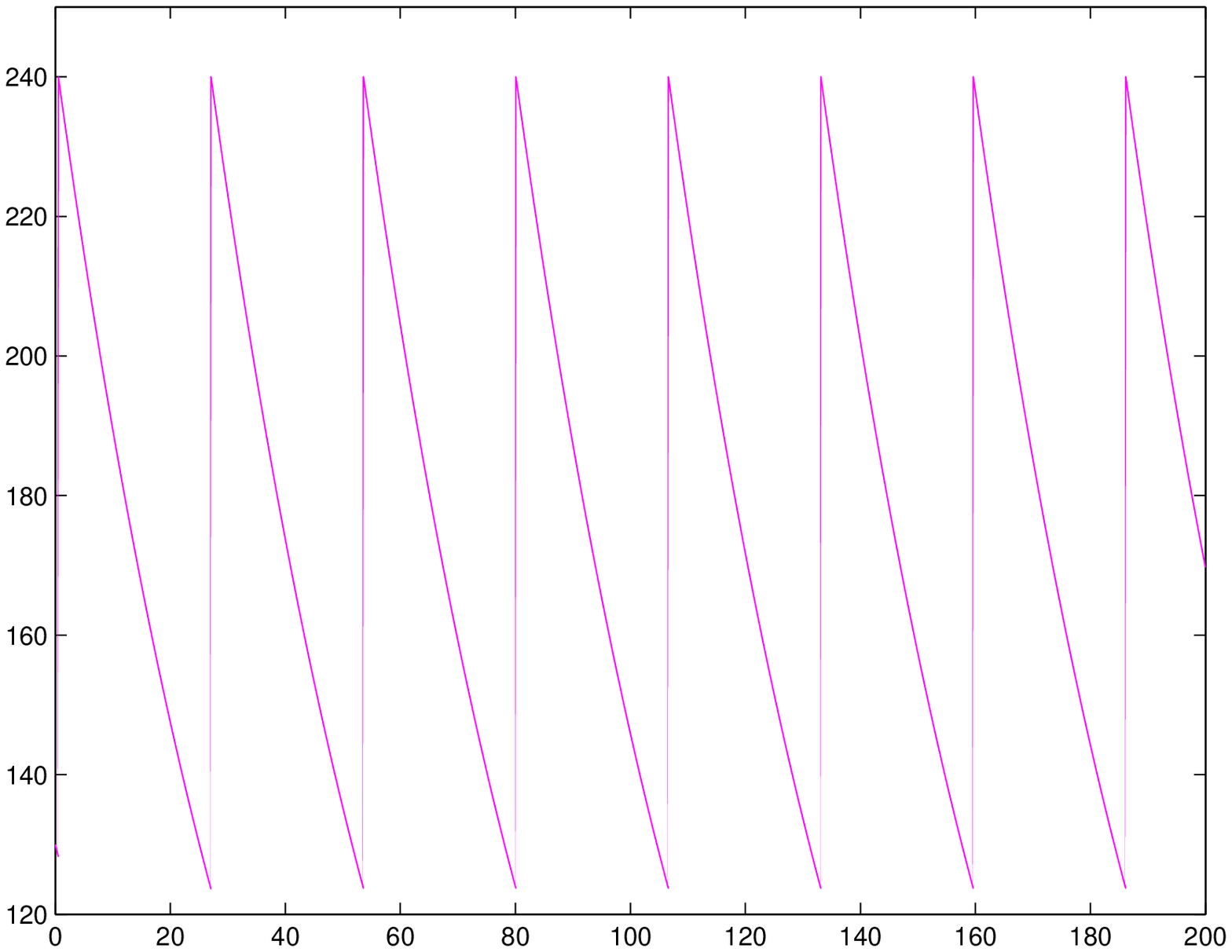,height=40mm,width=110mm}
\caption{$k$-WTA computation result of Example~\ref{ex:kwta} with $n=10$ and $k=3$.
The plots are the time developments of 
(a) $v_i$ of the neurons with the three largest inputs;
(b) $u_i$ of the neurons with the three largest inputs; 
(c) $v_i$ of the other seven neurons;
(d) $u_i$ of the other seven neurons; 
(e) global inhibition $z$.
The computation is completed in less than two periods.}
\label{fig:k_normal}
\end{center}
\end{figure}

\Example{}{The result is illustrated in simulation in
Figure~\ref{fig:k_normal}, with $n=10$ and $k=3$.  The parameters of
the FN neurons are set as $\ \alpha = 5.32, \beta = 3, \gamma = 0.1\
$, with spiking threshold $\ v_0 = 5\ $. The parameters of the local
inhibition are $\ u_0 = 160, k_u = 100\ $.  The inputs $I_i$ are
chosen randomly from $20$ to $125$.  The parameters of the global
neuron are $\ z_0 = 240, k_c = 100, k_d = 1/40\ $. All initial
conditions are chosen arbitrarily.  The three spiking neurons after
the first charging of the global neuron are those with the three
largest inputs.  

Note that the output frequency is determined mainly by the global
neuron's dynamics and the value of the $k^{\rm th}$ largest input. It
can be increased by increasing the global neuron's discharging rate
after the first winner spikes so as to facilitate the other
winners' spiking.}{kwta}

\begin{figure}[h]
\begin{center}
\epsfig{figure=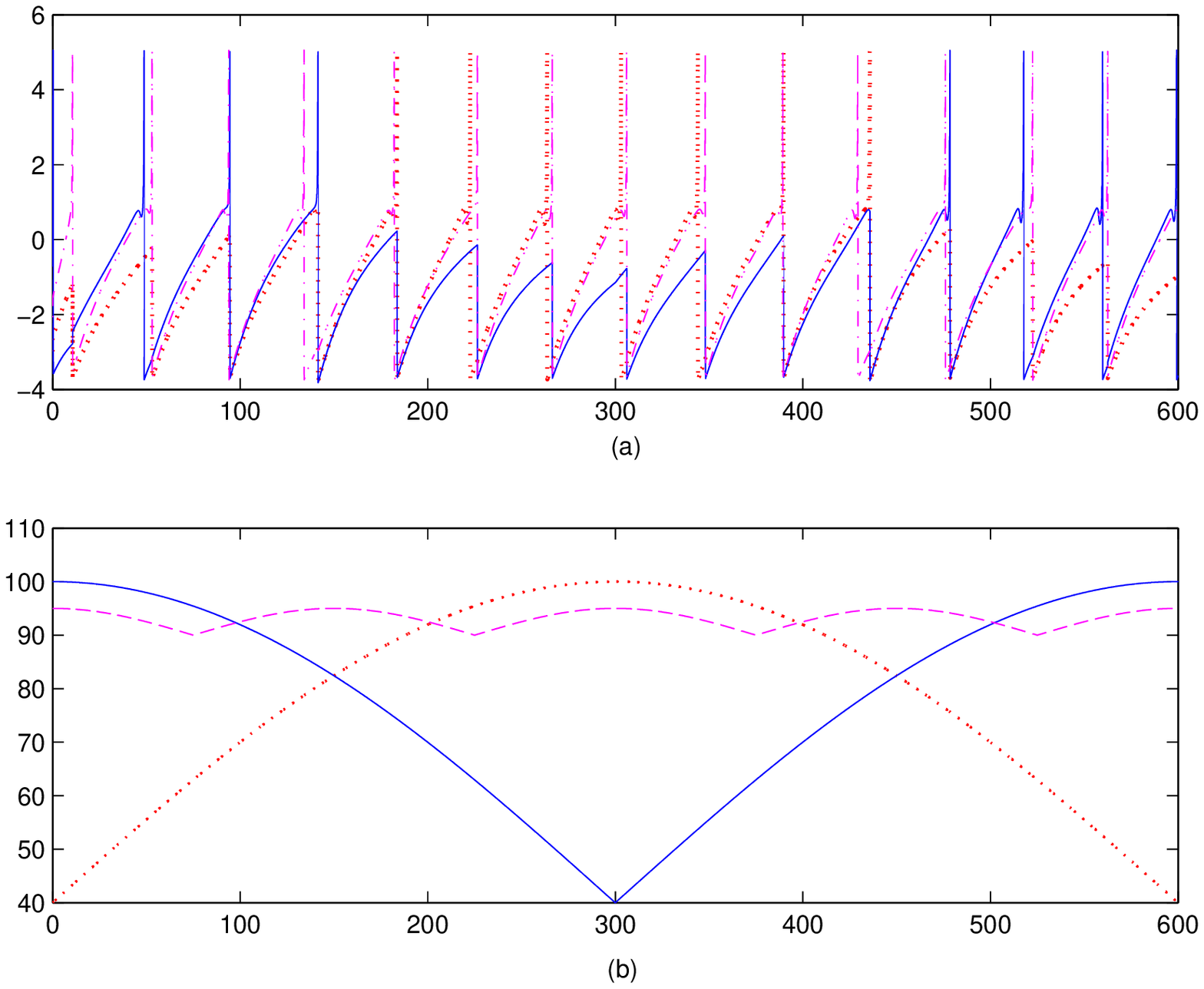,height=60mm,width=110mm}
\caption{$k$-WTA computation result of Example~\ref{ex:varying} with $n=3$ and $k=2$.
The inputs are not constant. The spiking neurons always track the two largest inputs.
The plots are (a).states $v_i$ versus time; (b).inputs $I_i$ versus time. Note that
each $v_i$ in plot (a) corresponds to the $I_i$ in plot (b) with the same line type
(solid, dashed or dotted).}
\label{fig:varying}
\end{center}
\end{figure}

\Example{}{Figure~\ref{fig:varying} illustrates a simulation result
with $n=3$ and $k=2$. The parameters are the same as those in
Example~\ref{ex:kwta}.  The inputs keep varying and switch winning
positions several times. The spiking neurons always track the two
largest inputs.}{varying}

\section{Soft-Winner-Take-All} \label{sec:soft}
Soft-WTA~\cite{maass00-1} (or softmax) is another variation of WTA
computation, where the outputs reflect the rank of all inputs
according to their size. Although soft-WTA is a very powerful
primitive~\cite{maass00-1, maass00-2} in that it can be used to
compute any continuous function, its ``neural'' implementation is
complex. Recently, \cite{yuille02} studied soft-WTA as an optimization
problem not based on a biologically plausible mechanism;
\cite{indiveri} presented a hardware model of selective visual
attention which lets the attention switch between the selected inputs,
but whose switching order does not completely reflect the input
ranks. In this section, we develop a simple neural network which
computes soft-WTA very fast and generates spiking outputs rank-ordered
by their inputs.

Let $k=n\ $ in the $k$-WTA network described in Section~\ref{sec:kwta}.
Then we get a pre-ordered spiking sequence in each stable period, since
all the FN neurons enter the oscillation region and then spike
rank-ordered by their inputs. However, such an $n^{\rm th}$ arrival
moment may not be measurable if the number of inputs $n$ is unknown or
time-varying. To avoid this problem and make the solution more
general, we let the charging mode of the global inhibitory neuron
start only if the inhibition $z$ is lower than a given bound
$z_{low}$. The spikings of all the FN neurons in the network are
guaranteed by the condition that
$$
z_{low} \ < \ I_l + I_{min} 
$$
where $I_l$ is the lower bound of the oscillation region of the 
FN model and $I_{min}$ is the minimum input value.

\begin{figure}[h]
\begin{center}
\epsfig{figure=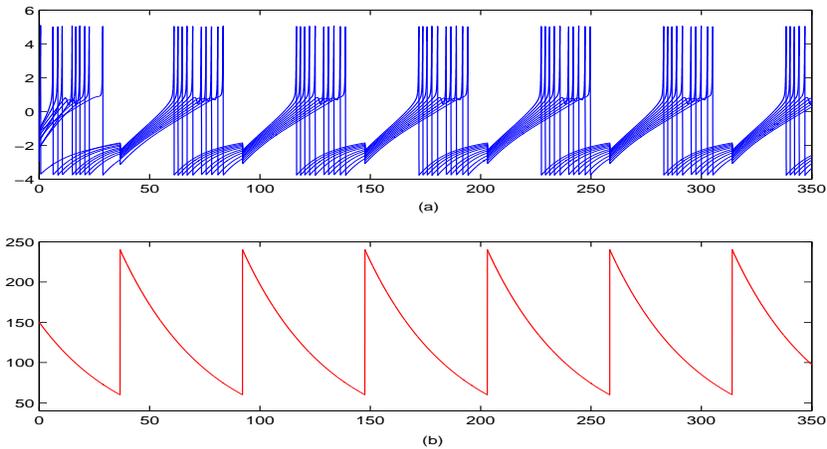,height=60mm,width=110mm}
\caption{Soft-WTA computation result of Example~\ref{ex:soft} with
$n=10$.  The plots are (a).states $v_i$ versus time; (b).global
inhibition $z$ versus time.  The initial conditions are chosen
arbitrarily and the computation is completed in the second period.}
\label{fig:soft}
\end{center}
\end{figure}

\Example{}{Figure~\ref{fig:soft} illustrates the result in simulation
with $n=10$.  The parameters are the same as those in
Example~\ref{ex:kwta}.  The inputs $I_i$ are distributed uniformly
between $80$ and $120$. The inhibition lower bound is $\ z_{low} = 60\
$.  Initial conditions are chosen arbitrarily.  The computation is
completed in the second period, during and after which the spiking
times of the FN neurons are ranked by their inputs.}{soft}

The simple soft-WTA network presented above inherits the main
computational advantages of our WTA and $k$-WTA networks. Initial
conditions can be arbitrary, the computation is completed in at most
two periods, network complexity is linear and neurons can be added or
removed at any time. We expect it to be effective in many applications
such as selective attention, associative memory and competitive
learning, and also to provide an efficient desynchronization mechanism
for perceptional binding~\cite{gray, singer, malsburg95, deliang02}.

\section{Concluding Remarks} \label{sec:conclusion}
Basic neural computations such as winner-take-all,
$k$-winners-take-all, soft-winner-take-all, and coincidence detection
can all be implemented using a common architecture and biological
plausible neuron models. Fast and robust convergence is guaranteed,
and time-varying inputs can be tracked. Further research will study
models of higher level brain functions, such as perception, based on
the WTA networks and on general nonlinear synchronization mechanisms
derived in~\cite{slotine03, wei03-2}.

\vspace{1.5em}

\noindent {\large{\bf Acknowledgments:}} This work was supported in
part by a grant from the National Institutes of Health. The authors
benefited from stimulating discussions with Matthew Tresch.

%
%
\renewcommand{\baselinestretch}{1.}

\end{document}